\documentclass[pdflatex,sn-nature,oneside]{sn-jnl}

\usepackage{graphicx}
\usepackage{multirow}
\usepackage{amsmath,amssymb,amsfonts}
\usepackage{amsthm}
\usepackage{mathrsfs}
\usepackage[title]{appendix}
\usepackage{xcolor}
\usepackage{textcomp}
\usepackage{manyfoot}
\usepackage{booktabs}
\usepackage{algorithm}
\usepackage{algorithmicx}
\usepackage{algpseudocode}
\usepackage{listings}

\unnumbered

\theoremstyle{thmstyleone}

\theoremstyle{thmstyletwo}

\theoremstyle{thmstylethree}

\begin{document}

\title[Scalable photoinduced molecular dynamics]
{\fontsize{17}{21}\selectfont\bfseries
Scalable photoexcitation-induced molecular dynamics with machine-learned Hamiltonians}

\author[1,2]{\fnm{Leyu} \sur{Cai}}
\author[1,2]{\fnm{Yunzhe} \sur{Jia}}
\author[1,2]{\fnm{Daqiang} \sur{Chen}}
\author*[1,2,3]{\fnm{Sheng} \sur{Meng}}\email{smeng@iphy.ac.cn}

\affil[1]{\orgdiv{Beijing National Laboratory for Condensed Matter Physics and Institute of Physics}, 
\orgname{Chinese Academy of Sciences}, 
\orgaddress{\city{Beijing}, \postcode{100190}, \country{China}}}

\affil[2]{\orgdiv{School of Physical Sciences}, 
\orgname{University of Chinese Academy of Sciences}, 
\orgaddress{\city{Beijing}, \postcode{100190}, \country{China}}}

\affil[3]{\orgname{Songshan Lake Materials Laboratory}, 
\orgaddress{\city{Dongguan}, \state{Guangdong}, \postcode{523808}, \country{China}}}


\abstract{
% < 150 words for ncs

\textbf{Ultrafast photoexcitation offers a controllable route to steer structural dynamics in solids, yet predicting how nonequilibrium electronic excitation drives lattice motion across extended spatial and temporal scales remains a major computational challenge. Here we introduce   time-dependent ab-initio propagation with electronic machine learning (TDAP-eML), a framework that explicitly incorporates electronic evolution into scalable simulations of photoexcitation-induced lattice dynamics. By integrating machine-learned electronic structure with atomistic propagation, TDAP-eML describes how photoexcitation reshapes the evolving energy landscapes and forces governing structural motion. Across representative examples including silicon and FeSe, the framework reproduces key photoexcited lattice responses obtained from first-principles time-dependent density functional theory calculations and captures coherent phonon dynamics together with their dependence on excitation conditions. Its computational advantage increases with system size, reaching nearly three orders of magnitude reduction in computational cost for the large systems examined. TDAP-eML thus establishes a scalable framework for coupled electronic and lattice evolution, linking nonequilibrium excitation to photoinduced forces, predictive structural dynamics, and experimentally accessible observables.}
}

%%================================%%
%% Sample for structured abstract %%
%%================================%%



\maketitle

\vspace{0.4em}
%\section{Introduction}\label{sec1}

Ultrafast photoexcitation can drive solids into nonequilibrium states and induce pronounced lattice dynamics on femtosecond-to-picosecond timescales~\cite{sidiropoulos2021probing,basov2017towards,delatorre2021nonthermal}. These photoexcited responses include coherent phonon oscillations~\cite{garrett1996coherent}, transient structural distortions~\cite{fritz2007ultrafast} and photoinduced phase transitions~\cite{cavalleri2001femtosecond,wall2012ultrafast}, with their characters and magnitudes depending sensitively on pump laser polarization, fluence and other photoexcitation conditions~\cite{misochko2015polarization,wall2012ultrafast}. At the microscopic level, photoexcitation redistributes electronic population, modifying chemical bonding and electron--phonon interactions, and thereby reshaping the forces acting on the lattice~\cite{lian2020ultrafast,liu2020microscopic}. The central challenge is therefore to follow the full causal chain from photoexcited electronic redistribution to modified atomic forces and, ultimately, collective structural motions~\cite{zeiger1992theory,shinohara2010coherent}.

First-principles time-dependent approaches such as  time-dependent density functional theory (TDDFT) provide a direct microscopic description of photoexcitation-induced lattice dynamics, but their heavy computational cost severely restricts the accessible system sizes and timescales~\cite{yamada2019multiscale,tong2021toward,guan2023coherent}. Recent advances in machine learning (ML), particularly in machine-learned electronic Hamiltonians, offer an efficient, geometry-dependent representation of electronic structure, opening new possibilities for incorporating information on electronic evolution into large-scale atomistic simulations~\cite{schutt2019unifying,li2022deeph,zhang2022equivariant,zhong2023transferable}. Recent studies have also explored the application of machine-learned
Hamiltonians to nonadiabatic molecular dynamics within the fewest-switches
surface-hopping framework
~\cite{cao2025largescale,zhang2025n2amd,zhang2026nonadiabatic}. A distinct challenge that remains in simulating light-driven materials is to determine how the evolving photoexcited electronic state modifies the forces that drive collective lattice motion. Addressing this problem for realistic large-scale materials simulations requires a computational framework that connects 
electronic evolution directly to excited-state structural dynamics.

Here we introduce the time-dependent ab-initio propagation scheme with electronic machine learning (TDAP-eML), a framework for photoexcitation-induced lattice dynamics that explicitly incorporates information on electronic excitation and population evolution. TDAP-eML incorporates photoinduced electronic redistribution to drive lattice dynamics by combining the nonequilibrium electronic occupation with a differentiable, structure-dependent ML electronic Hamiltonian to construct occupation-dependent effective energy landscapes and atomic forces, while the machine-learning force field (MLFF) provides the reference lattice energetics. This framework is explicitly tested against two prototype photoexcited systems. In silicon, we show how 
photoexcited electronic redistribution reshapes the lattice energy landscape and driving forces, giving rise to coherent phonon dynamics with pronounced polarization and fluence dependence together with transient optical signatures under experimentally relevant excitation conditions. The resulting photoexcited PESs and dominant lattice dynamics are quantitatively validated against real-time TDDFT simulations~\cite{runge1984density,meng2008real,lian2018photoexcitation} and the predicted transient optical spectrum agrees well with the experimental measurements. Applications to FeSe and larger simulation cells further demonstrate the general applicability of TDAP-eML across distinct material environments and computational advantages that increase with system size. TDAP-eML thus establishes a scalable framework linking evolving electronic excitation to photoinduced atomic forces, coherent structural motion, and experimentally accessible responses.

\section{Results}\label{sec2}
\subsection{Theory and framework of TDAP-eML}

\begin{figure}[!t]
  \centering    
\includegraphics[width=1\textwidth]{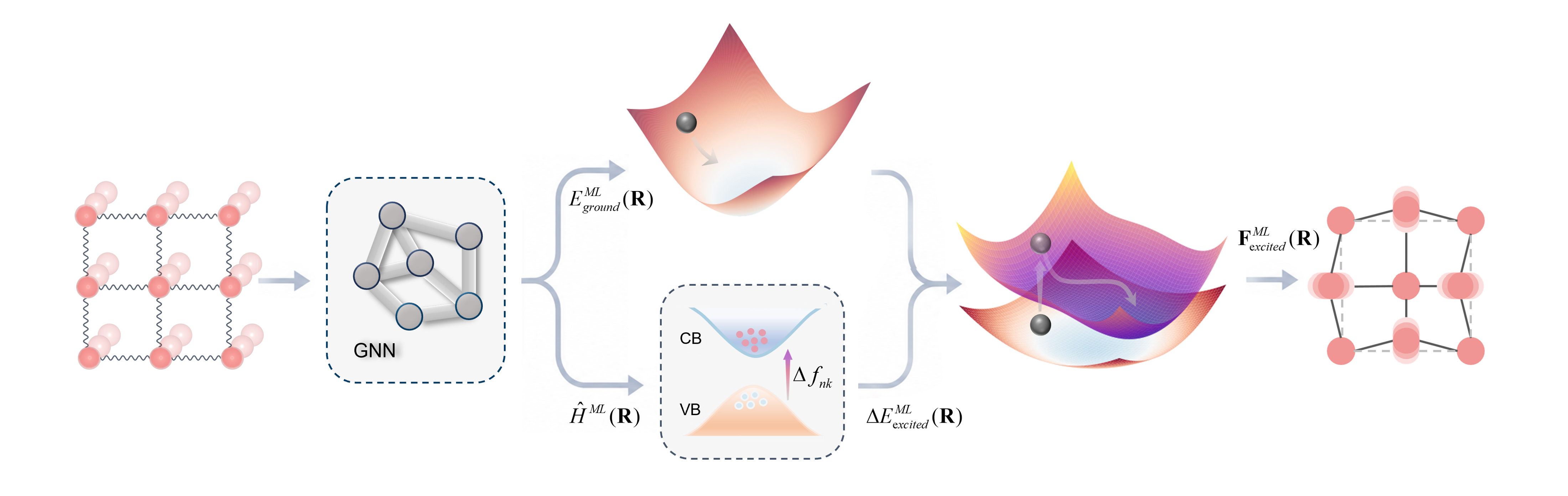}
\caption{
\textbf{Schematic illustration of the TDAP-eML framework.}
An atomic configuration $\mathcal{R}$ is evaluated by complementary MLFF and ML electronic Hamiltonian branches, which together construct the effective excited-state PES $E_e^{\mathrm{ML}}(\mathcal{R})$ and the corresponding atomic forces. The ML Hamiltonian provides the geometry-dependent electronic structure, while the photoinduced occupation change $\Delta f_{n\mathbf{k}}$ introduces the excitation dependence of the PES and atomic forces.
}
    \label{fig1}
\end{figure}

Figure~\ref{fig1} summarizes the TDAP-eML framework, which connects photoexcited electronic populations to lattice dynamics through a geometry-dependent electronic Hamiltonian. The underlying physical picture is motivated by the conventional displacive excitation of coherent phonons (DECP) mechanism, in which laser-induced changes in electronic occupations generate a driving force on the lattice through electron--phonon coupling~\cite{zeiger1992theory,garrett1996coherent}. Within a linear electron--phonon coupling picture, the photoinduced force for phonon mode $\nu$ can be expressed schematically as
$\Delta F_{\nu}^{e}\propto-\sum_{n,\mathbf{k}}w_{\mathbf{k}}g_{n\mathbf{k}}^\nu\Delta f_{n\mathbf{k}}$,
where $w_{\mathbf{k}}$ is the $\mathbf{k}$-point weight, $g_{n\mathbf{k}}^\nu$ denotes the coupling of the $n$th electronic state at $\mathbf{k}$ to this phonon mode, and $\Delta f_{n\mathbf{k}}=f_{n\mathbf{k}}-f^{g}_{n\mathbf{k}}$ denotes the laser-induced change in the occupation $f_{n\mathbf{k}}$ of state $(n,\mathbf{k})$ relative to its ground-state value $f^{g}_{n\mathbf{k}}$~\cite{wang2025directional,emeis2025coherent}. Recent studies have shown that ML electronic Hamiltonians can accurately capture electron--phonon coupling and its dependence on atomic structure~\cite{zhong2024accelerating}, motivating their use as an electronic representation for describing how photoexcitation drives collective lattice motion.

To this end, TDAP-eML constructs an occupation-dependent effective PES for the photoexcited lattice and derives the corresponding atomic forces. For an atomic configuration $\mathcal{R}=\{\mathbf{R}_i\}$, the reference ground-state PES provided by the MLFF is combined with an electronic contribution obtained from the geometry-dependent ML electronic Hamiltonian $\hat{H}_{\mathbf{k}}(\mathcal{R})$. The excitation condition enters through the nonequilibrium occupation variation $\Delta f_{n\mathbf{k}}$, giving
\begin{equation}
\Delta E_e(\mathcal{R})
=
\sum_{n,\mathbf{k}}
w_{\mathbf{k}}
\Delta f_{n\mathbf{k}}
\varepsilon_{n\mathbf{k}}(\mathcal{R}),
\label{eq:excited_energy_correction}
\end{equation}
where $\varepsilon_{n\mathbf{k}}(\mathcal{R})$ is the corresponding instantaneous electronic eigenvalue of the ML Hamiltonian. Together with the reference ground-state PES $E_g(\mathcal{R})$, this contribution defines the effective PES governing the photoexcited lattice dynamics,
\begin{equation}
E_e(\mathcal{R})
=
E_g(\mathcal{R})
+
\Delta E_e(\mathcal{R}).
\label{eq:excited_pes}
\end{equation}

Physically, this construction makes the lattice energy landscape dependent on the photoexcited electronic occupation, through which the effect of electronic excitation and evolution is incorporated into the forces governing the subsequent atomic motion. 
The resulting $E_e(\mathcal{R})$ therefore represents an occupation-constrained effective excited-state PES rather than a fully self-consistent excited-state total-energy functional. Self-consistent photocarrier-induced renormalization of the electronic Hamiltonian and more general nonadiabatic electron--lattice effects are not included in the present formulation~\cite{tong2021toward}.

The atomic forces governing the photoexcited lattice dynamics are obtained directly from the gradient of the effective PES in Eq.~\eqref{eq:excited_pes}. For the $i$th atom,
\begin{equation} 
\mathbf{F}_i^e = -\nabla_i E_e(\mathcal{R}) = \mathbf{F}_i^g + \Delta\mathbf{F}_i^e, 
\label{eq:force_decomposition} 
\end{equation} 
where $\mathbf{F}_i^g=-\nabla_iE_g(\mathcal{R})$ provides the reference force contribution and $\nabla_i\equiv\partial/\partial\mathbf{R}_i$. The photoinduced force correction is then given by
\begin{equation}
\Delta\mathbf{F}_i^e = -\sum_{n,\mathbf{k}} w_{\mathbf{k}} \Delta f_{n\mathbf{k}} \nabla_i\varepsilon_{n\mathbf{k}}(\mathcal{R}). 
\label{eq:excited_force_correction} 
\end{equation}
Equation~\eqref{eq:excited_force_correction} shows that the photoinduced occupation changes weight the gradients of the individual electronic eigenvalues, thereby providing a direct microscopic connection between nonequilibrium electronic redistribution and the forces governing lattice dynamics.

Evaluation of the occupation-dependent force contribution requires the derivatives of the electronic eigenvalues with respect to the atomic coordinates. For a non-degenerate state in the localized nonorthogonal basis employed by the ML Hamiltonian, the eigenvalue derivative is given by~\cite{pulay1969abinitio}
\begin{equation} 
\nabla_i \varepsilon_{n\mathbf{k}} = \mathbf{c}_{n\mathbf{k}}^{\dagger} \left[ \nabla_i \hat{H}_{\mathbf{k}}(\mathcal{R}) - \varepsilon_{n\mathbf{k}} \nabla_i \hat{S}_{\mathbf{k}}(\mathcal{R}) \right] \mathbf{c}_{n\mathbf{k}},
\label{eq:hf_derivative}
\end{equation}
where $\mathbf{c}_{n\mathbf{k}}=\{ c_{\mu n\mathbf{k}} \} $ denotes a column of the expansion coefficients of the $n$th
instantaneous electronic eigenstate $|\psi_{n\mathbf{k}} \rangle$ in the localized basis $|\phi_{\mu\mathbf{k}}\rangle$, and satisfies the
generalized eigenvalue problem
$\hat{H}_{\mathbf{k}}\mathbf{c}_{n\mathbf{k}}
=
\varepsilon_{n\mathbf{k}}
\hat{S}_{\mathbf{k}}\mathbf{c}_{n\mathbf{k}}$
with $\hat{S}_{\mathbf{k}}$ being the corresponding overlap matrix.
The Hamiltonian derivative $\nabla_i\hat{H}_{\mathbf{k}}$ can be evaluated
efficiently through automatic neural network differentiation of the ML Hamiltonian with
respect to the atomic coordinates. The term involving the overlap derivative,
$-\varepsilon_{n\mathbf{k}}\nabla_i\hat{S}_{\mathbf{k}}$ arises from the
geometry dependence of the nonorthogonal basis and gives the Pulay contribution
to the photoinduced forces in Eq.~\eqref{eq:excited_force_correction}
(see Supplementary Note~3 for details).

To couple the electronic evolution to the atomic motion, we further consider
the time dependence of $\Delta f_{n\mathbf{k}}$. Its initial post-pulse value
is obtained from a reference TDDFT calculation~\cite{meng2008real,lian2018photoexcitation},
while its subsequent evolution along the atomic trajectory is governed by the
time-dependent Schr\"odinger equation (TDSE),
\begin{equation}
i\hbar
\frac{\partial}{\partial t}
|\Psi_{\mathbf{k}}(t)\rangle
=
\hat{H}_{\mathbf{k}}[\mathcal{R}(t)]
|\Psi_{\mathbf{k}}(t)\rangle .
\label{eq:tdse}
\end{equation}
Expanding the time-dependent electronic state in the instantaneous eigenstates yields,
$
|\Psi_{\mathbf{k}}(t)\rangle
=
\sum_n a_{n\mathbf{k}}(t)|\psi_{n\mathbf{k}}(t)\rangle
=
\sum_{n,\mu}
a_{n\mathbf{k}}(t)c_{\mu n\mathbf{k}}(t)
|\phi_{\mu\mathbf{k}}(t)\rangle ,
$
where $|\psi_{n\mathbf{k}}(t)\rangle$ is the instantaneous eigenstate of
$\hat{H}_{\mathbf{k}}[\mathcal{R}(t)]$  
and $|\phi_{\mu\mathbf{k}}(t)\rangle$ is the localized basis function.
Accordingly, $a_{n\mathbf{k}}(t)$ gives the time-dependent amplitude in the adiabatic basis of instantaneous eigenstates $|\psi_{n\mathbf{k}} \rangle$.

The TDSE then yields
\begin{equation}
i\hbar
\frac{\partial a_{n\mathbf{k}}(t)}{\partial t}
=
\varepsilon_{n\mathbf{k}}(t)a_{n\mathbf{k}}(t)
-
i\hbar
\sum_m
\left\langle
\psi_{n\mathbf{k}}(t)
\middle|
\frac{\partial}{\partial t}
\psi_{m\mathbf{k}}(t)
\right\rangle
a_{m\mathbf{k}}(t).
\label{eq:a_evolution}
\end{equation}
For a sufficiently small time step $\Delta t$, the integral in the second term on the right-hand side can be represented
through the overlap between instantaneous states at successive atomic
configurations. To  first order in $\Delta t$,
\begin{equation}
a_{n\mathbf{k}}(t+\Delta t)
=
\sum_m
\left\langle
\psi_{n\mathbf{k}}(t+\Delta t)
\middle|
\psi_{m\mathbf{k}}(t)
\right\rangle
a_{m\mathbf{k}}(t)
-
\frac{i\Delta t}{\hbar}
\varepsilon_{n\mathbf{k}}(t)
a_{n\mathbf{k}}(t)
+
\mathcal{O}(\Delta t^2).
\label{eq:a_discrete}
\end{equation}
The finite-step representation directly connects the electronic evolution to the structural change between successive MD steps.
The corresponding electronic population is
$f_{n\mathbf{k}}(t)=|a_{n\mathbf{k}}(t)|^2$, yielding the time-dependent occupation change
$\Delta f_{n\mathbf{k}}(t)=f_{n\mathbf{k}}(t)-f_{n\mathbf{k}}^{\mathrm g}$. The updated occupations determine the effective PES and atomic
forces through Eqs.~\eqref{eq:excited_pes} and
\eqref{eq:excited_force_correction}. These photoinduced forces propagate the lattice in the excited state, and the resulting structural change updates the ML electronic Hamiltonian and the instantaneous electronic states at the next MD step, thereby closing the feedback loop between electronic evolution and atomic motion. TDAP-eML realizes this coupled propagation efficiently by replacing repeated first-principles evaluations along the MD trajectory with ML inference and automatic differentiation of the geometry-dependent electronic Hamiltonian, which provide the instantaneous electronic structure and the atomic-coordinate derivatives required to construct the photoinduced force correction.

\subsection{Photoinduced reshaping of lattice PES in silicon}

\begin{figure}[!t]
  \centering  
  \includegraphics[width=1\textwidth]{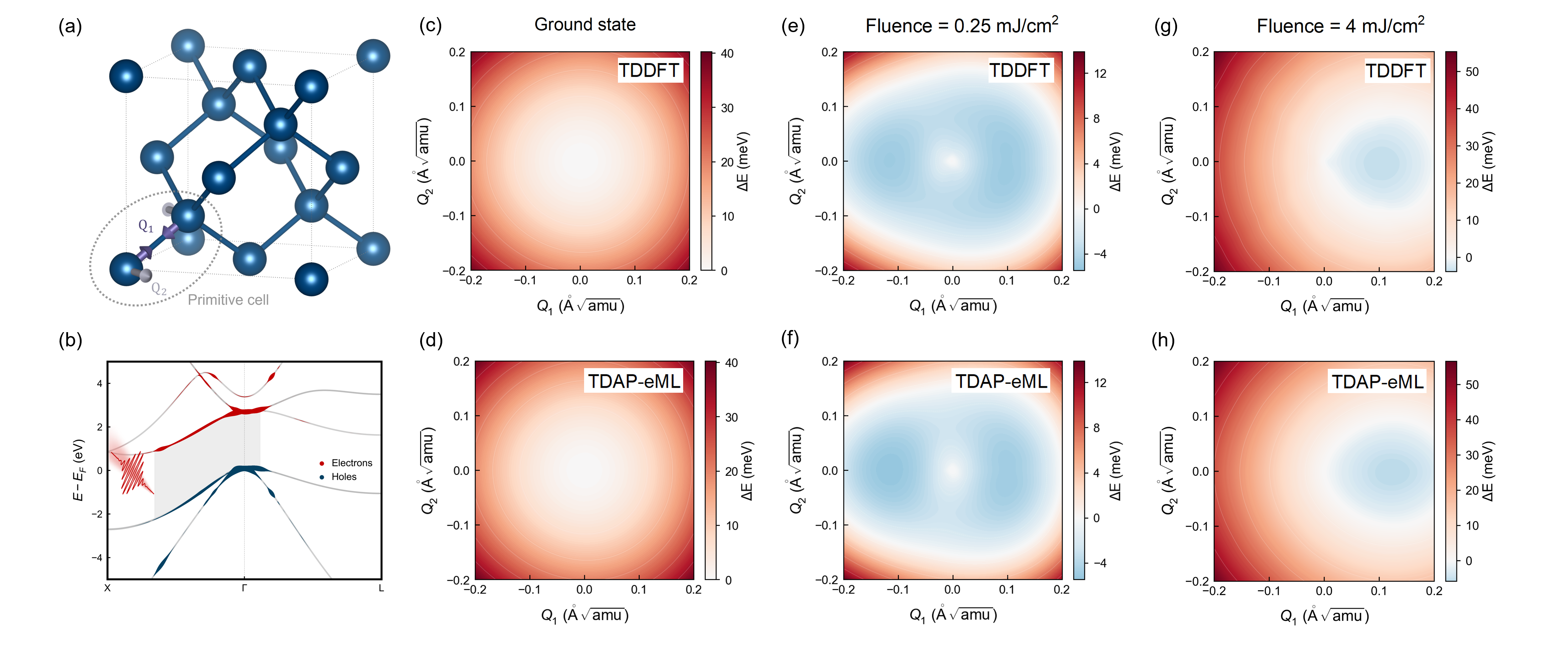}
\caption{
\textbf{Photoexcited PES for lattice motion in silicon.}
\textbf{a,} Definition of the optical phonon coordinates shown in silicon, with $Q_1\parallel[111]$ and $Q_2\parallel[1\bar{1}0]$.
\textbf{b,} Representative photoexcited changes in electron occupation $\Delta f_{n\mathbf{k}}$ obtained from TDDFT under a $[111]$-polarized laser field with a fluence of $0.25~\mathrm{mJ\,cm^{-2}}$, shown along the $X$--$\Gamma$--$L$ path.
\textbf{c--h,} Ground-state and photoexcited PESs in the $Q_1$--$Q_2$ plane calculated using TDDFT and TDAP-eML, respectively. The lower- and higher-fluence photoexcited PESs correspond to a fluence of $0.25~\mathrm{mJ\,cm^{-2}}$ and $4~\mathrm{mJ\,cm^{-2}}$, respectively. Energies are referenced to the undistorted ground-state equilibrium structure.
}
  \label{fig2}
\end{figure}

To demonstrate the capability of TDAP-eML for describing photoexcited lattice dynamics, we first apply it to silicon, a prototypical system with well-established coherent optical phonon responses~\cite{hase2003birth,hase2013coherent}. Its well-characterized photoexcited responses provide a natural platform for examining how photoexcitation reshapes the energy landscape of the lattice and drives coherent structural motion. The triply degenerate optical phonons at the $\Gamma$ point are represented by three orthogonal normal coordinates, $Q_1$, $Q_2$ and $Q_3$, as defined in Supplementary Note~4 and chosen along $[111]$, $[1\bar{1}0]$ and $[11\bar{2}]$, respectively [Fig.~\ref{fig2}(a)]. A Gaussian pump pulse polarized along $[111]$ generates the nonequilibrium electron occupation used to construct the photoexcited PES, with a representative  occupation change $\Delta f_{n\mathbf{k}}$ at $0.25~\mathrm{mJ\,cm^{-2}}$ shown in Fig.~\ref{fig2}(b).

Figure~\ref{fig2}(c)--(h) shows the ground-state PES and the photoexcited PESs at two representative pump fluences in the $Q_1$--$Q_2$ plane, calculated using both TDDFT and TDAP-eML. In the ground state, the energy minimum lies at the undistorted equilibrium structure. Photoexcitation reshapes the local energy landscape, with the most pronounced modification occurring along $Q_1$. At the higher fluence of $4~\mathrm{mJ\,cm^{-2}}$, the low-energy region is clearly displaced toward positive $Q_1$, indicating a pronounced photoinduced driving force along this phonon coordinate. Since $Q_1$ describes atomic motion along a $[111]$ Si--Si bond, the preferential response along $Q_1$ indicates that $[111]$-polarized laser excitation produces a displacive driving force primarily along the same bond direction.

The direct comparison with TDDFT provides a quantitative benchmark of the TDAP-eML photoexcited PESs. Across the full plotted regions in Fig.~\ref{fig2}(e--h), TDAP-eML yields mean absolute errors (MAEs) of $0.31~\mathrm{meV}$ and $0.84~\mathrm{meV}$ for the lower- and higher-fluence cases, respectively. A corresponding benchmark at the intermediate fluence of $1~\mathrm{mJ\,cm^{-2}}$ gives an MAE of $0.65~\mathrm{meV}$ (Supplementary Fig.~3). TDAP-eML thus reproduces both the overall energy variation across the PES and its photoexcitation-induced reshaping, with the dominant modification along $Q_1$, thereby providing a quantitative basis for the photoinduced lattice dynamics examined below.

\subsection{Coherent phonon dynamics on photoexcited PESs}

\begin{figure}[!t]
  \centering    
  \includegraphics[width=1\textwidth]{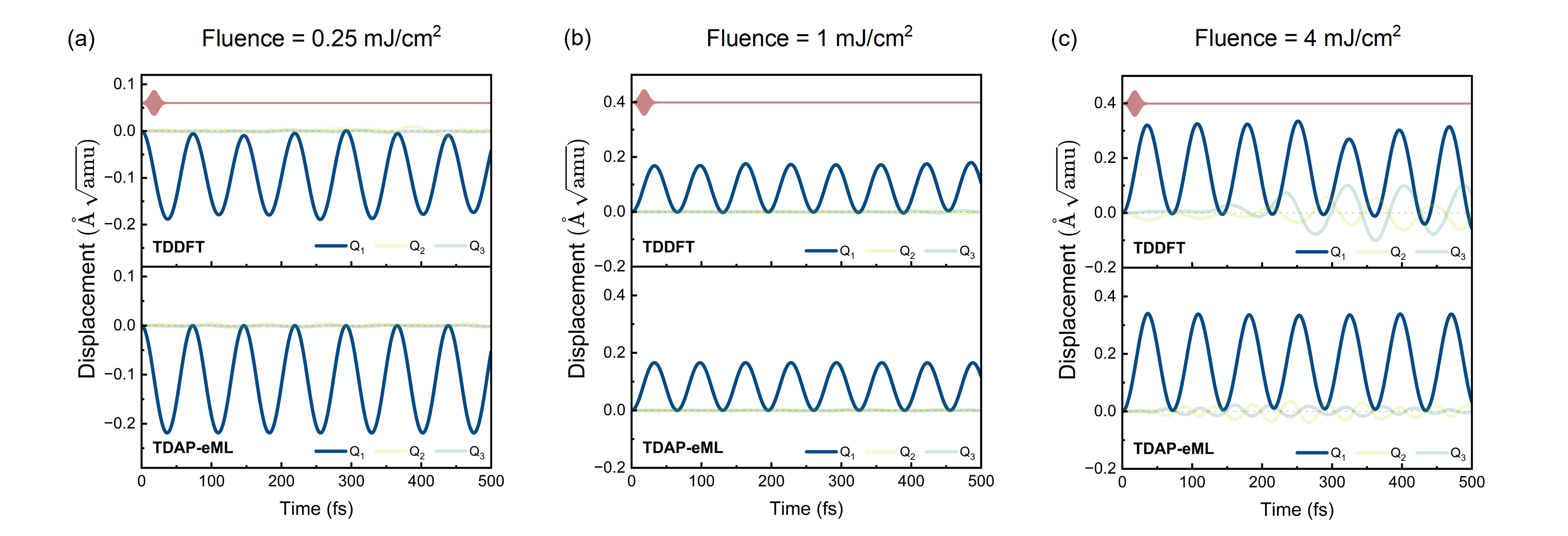}
\caption{
\textbf{Coherent phonon dynamics in silicon.}
\textbf{a--c,} Time evolution of the mass-weighted phonon coordinates $Q_1$, $Q_2$ and $Q_3$ at pump fluences of $0.25$, $1$, and $4~\mathrm{mJ\,cm^{-2}}$, respectively. The upper and lower subpanels show the corresponding TDDFT and TDAP-eML trajectories, respectively. The red profile in each upper subpanel shows the temporal profile of the laser electric field.
}
  \label{fig3}
\end{figure}

To assess how the photoexcited PESs translate into coherent lattice motion, Fig.~\ref{fig3} compares the phonon trajectories obtained from TDDFT and TDAP-eML  at three pump fluences. At lower fluence, the lattice response is dominated by $Q_1$, while  $Q_2$ and $Q_3$ remain nearly inactive. This behaviour is characteristic of a predominantly displacive excitation along the $[111]$ bond-oriented $Q_1$ coordinate, in line with the directional PES response identified in Fig.~\ref{fig2}.
At the intermediate fluence of $1~\mathrm{mJ\,cm^{-2}}$, the dynamics remain dominated by $Q_1$, with close agreement between TDAP-eML and TDDFT.
At the higher fluence, the $Q_1$ trajectory changes appreciably, while weak nonlinear phonon components also emerge along $Q_2$ and $Q_3$. Quantitative differences between TDAP-eML and TDDFT become more evident for these weaker components, which may partly reflect the increasing importance of photocarrier-induced electronic structure renormalization at stronger excitation. Despite these differences, the dominant dynamics in both calculations remain governed by $Q_1$, showing that TDAP-eML captures the principal photoinduced lattice response across all three excitation regimes. The predominance of the $Q_1$ mode further highlights the preferential driving of the phonon coordinate aligned with the pump-selected Si--Si bond direction.

\subsection{Polarization-dependent coherent lattice dynamics and transient optical response}

\begin{figure}[!t]
  \centering    
  \includegraphics[width=\textwidth]{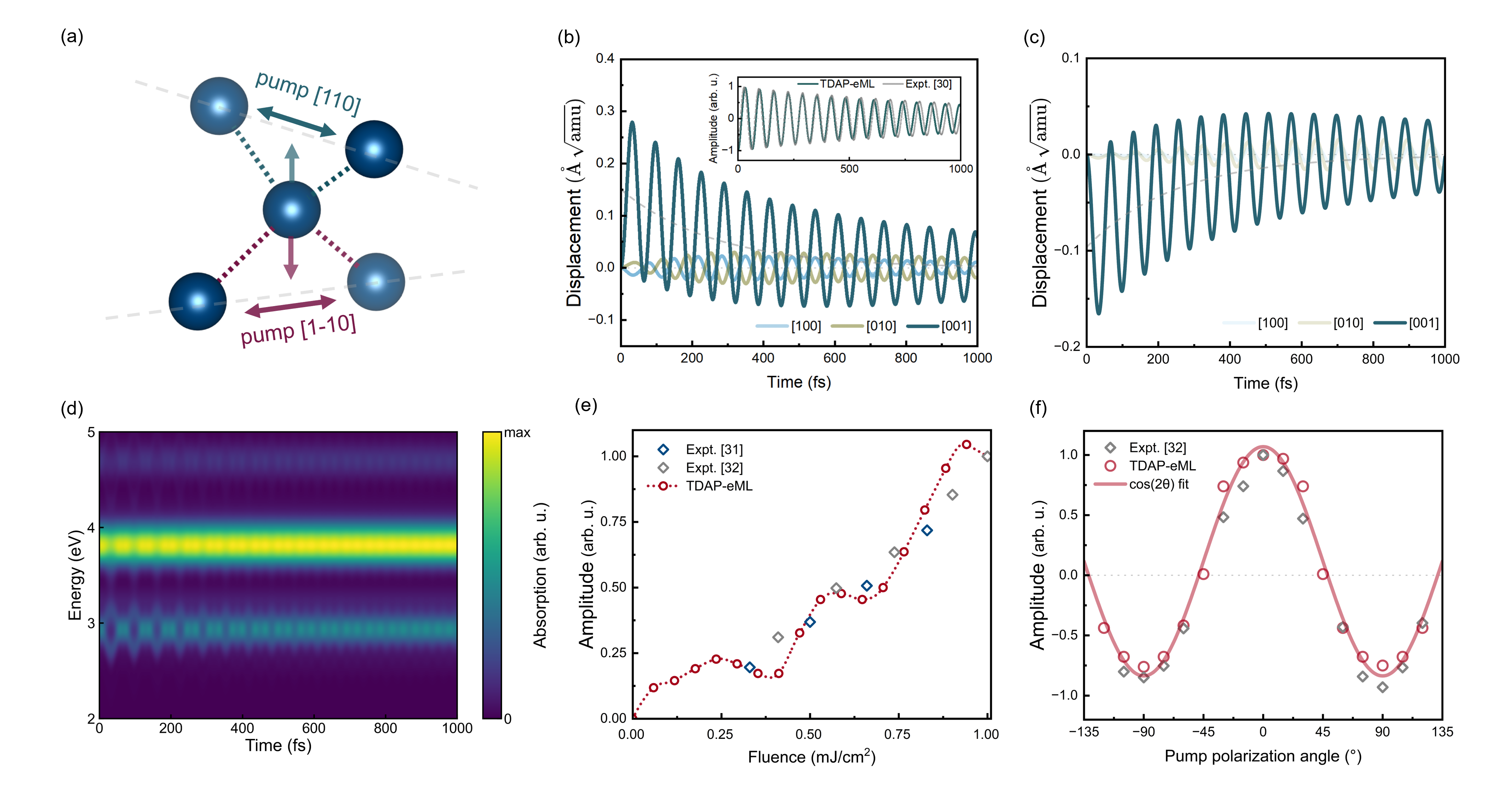}
  \caption{
  \textbf{Pump-dependent coherent phonon dynamics and transient optical response in silicon.}
  \textbf{a,} Si bonding geometry for pump polarizations along the in-plane $[110]$ and $[1\bar{1}0]$ directions, which generate opposite displacive forces along $[001]$.
  \textbf{b,c,} TDAP-eML displacement trajectories for $[110]$- and $[1\bar{1}0]$-polarized excitation, respectively. The inset in \textbf{b} compares the normalized TDAP-eML trajectory with experimental data from Ref.~\cite{hase2003birth}.
  \textbf{d,} Time-resolved absorption spectrum evaluated along the TDAP-eML trajectory.
  \textbf{e,} Coherent phonon amplitude as a function of pump fluence, compared with experimental data from Refs.~\cite{hase2013coherent,constantinescu2010carrier}.
  \textbf{f,} Coherent phonon amplitude as a function of pump polarization angle, compared with experimental data from Ref.~\cite{constantinescu2010carrier}. The solid line shows a $\cos(2\theta)$ fit.
  }
  \label{fig4}
\end{figure}

We next examine whether the directional photoinduced responses identified above can be translated into controllable coherent phonon motion. We consider bulk silicon with the pump polarization lying in the $(001)$ plane, focusing either on the orthogonal $[110]$ or $[1\bar{1}0]$ directions [Fig.~\ref{fig4}(a)].

For a pump photon energy of $\hbar\omega=2.6~\mathrm{eV}$, we propagate the longer-time lattice dynamics including carrier and lattice relaxation effects, as detailed in Methods. The $[110]$-polarized excitation produces a clear displacive coherent oscillation along $[001]$ [Fig.~\ref{fig4}(b)]. A damped displacive fit yields a phonon frequency of $15.57~\mathrm{THz}$ and a damping time of $1.18~\mathrm{ps}$, with the normalized trajectory reproducing nicely the characteristic oscillation period and decay observed experimentally~\cite{hase2003birth}.

Rotating the pump polarization by $90^\circ$ to $[1\bar{1}0]$ reverses the sign of the photoinduced displacement [Fig.~\ref{fig4}(c)]. Extending the polarization continuously within the $(001)$ plane, the coherent displacement follows an approximately $\cos(2\theta)$ angular dependence [Fig.~\ref{fig4}(f)], consistent with the experimentally observed symmetry~\cite{constantinescu2010carrier}. This continuous dependence connects the opposite responses at the two orthogonal polarizations within a common polarization-controlled lattice response. In addition, the coherent-phonon amplitude varies nonlinearly with pump fluence [Fig.~\ref{fig4}(e)], following a trend very similar to that observed experimentally~\cite{hase2013coherent,constantinescu2010carrier}.

The coherent lattice motion also modulates the electronic structure along the trajectory. The transient absorption spectrum evaluated from the geometry-dependent Hamiltonian [Fig.~\ref{fig4}(d)] exhibits periodic changes in both near-gap and higher-energy absorption features, providing an optical signature of the driven phonon motion. Together, these results demonstrate the ability of TDAP-eML to bridge microscopic photoinduced forces and experimentally observable coherent lattice dynamics, capturing their polarization and fluence dependence as well as the resulting optical signatures.

\subsection{Applicability beyond silicon and computational efficiency}

\begin{figure}[t]
  \centering    
  \includegraphics[width=1\textwidth]{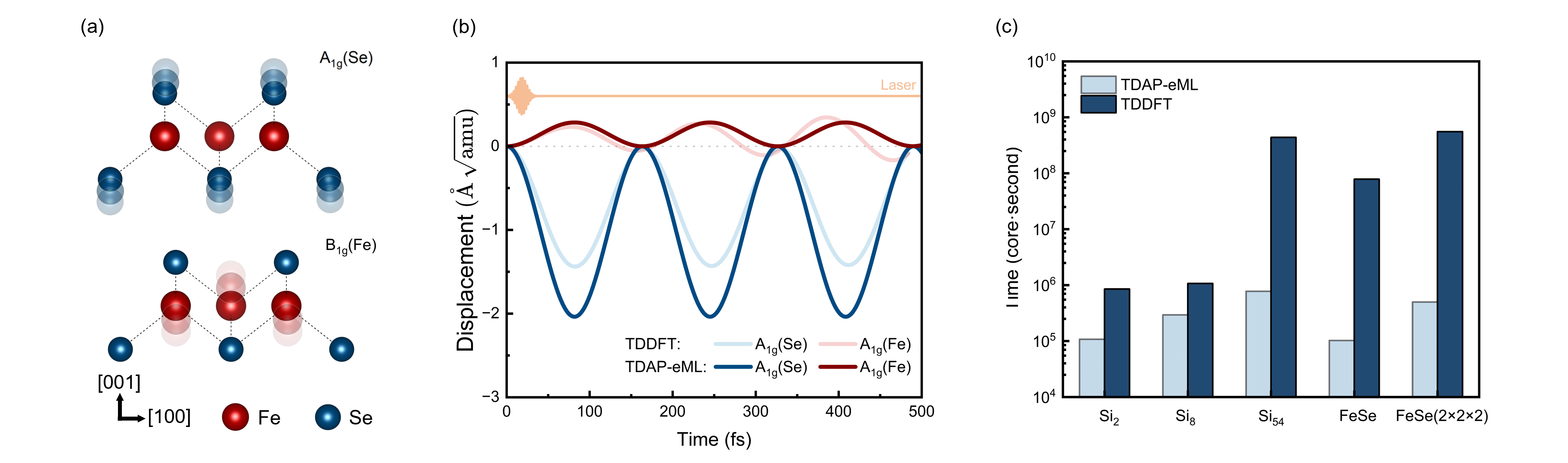}
\caption{
\textbf{Photoinduced lattice dynamics in FeSe and computational efficiency of the TDAP-eML scheme.}
\textbf{a,} Out-of-plane $A_{1g}(\mathrm{Se})$ and $B_{1g}(\mathrm{Fe})$ phonon modes in tetragonal FeSe, dominated by Se and Fe displacements, respectively.
\textbf{b,} Mode-resolved coherent phonon dynamics of the $A_{1g}(\mathrm{Se})$ and $B_{1g}(\mathrm{Fe})$ modes obtained from TDDFT and TDAP-eML.
\textbf{c,} Computational costs of TDDFT and TDAP-eML simulations for Si and FeSe cells of different sizes, reported in CPU core-seconds for a $1000~\mathrm{fs}$ trajectory of $10{,}000$ time steps.
}

  \label{fig5}
\end{figure}

We finally apply TDAP-eML to FeSe to assess its general applicability beyond a prototypical semiconductor. FeSe hosts multiple Fe $3d$-derived bands near the Fermi level and provides a more complex multi-orbital setting with strong sensitivity of the electronic structure to lattice geometry~\cite{gerber2017femtosecond,lazarevic2022evolution}. Figure~\ref{fig5}(a,b) compares the mode-resolved photoinduced lattice dynamics obtained from TDDFT and TDAP-eML simulations. In both calculations, the response is concentrated in the out-of-plane $A_{1g}(\mathrm{Se})$ and $B_{1g}(\mathrm{Fe})$ modes, while contributions from the other phonon coordinates are negligible. Despite quantitative differences in the oscillation amplitudes, TDAP-eML reproduces the dominant coherent modes and their temporal evolution obtained from TDDFT.

Figure~\ref{fig5}(c) further compares the computational cost for increasingly large simulation cells, including a $3\times3\times3$ Si supercell (Si$_{54}$) and a $2\times2\times2$ FeSe supercell. Across the systems considered, TDAP-eML remains substantially less expensive than TDDFT simulations, with the cost reduction reaching approximately three orders of magnitude for the larger and more complex systems. Together, the FeSe dynamical benchmark and the computational-cost comparison show that TDAP-eML captures the dominant photoinduced lattice responses in a more complex material setting, while offering a growing computational advantage as the system size increases.

\section{Discussion}

TDAP-eML establishes an integrated framework for photoexcitation-induced lattice dynamics that explicitly incorporates electronic population evolution. The geometry-dependent and differentiable ML Hamiltonian couples the evolving nonequilibrium electron occupation to structural responses of the electronic state, thereby continuously updating the effective excited-state PES and photoinduced forces along the atomic trajectory. This provides a microscopic connection between photoexcited electronic redistribution, electron--phonon coupling, and lattice motion, as illustrated in Si by the polarization- and fluence-dependent coherent responses along different lattice coordinates. TDAP-eML therefore enables both predictive simulations of photoinduced structural dynamics and a direct physical interpretation of how photoexcitation drives coherent lattice motion.

The force construction in Eq.~\eqref{eq:excited_force_correction} employs an occupation-constrained approximation, with the instantaneous $\Delta f_{n\mathbf{k}}(t)$ entering as time-dependent weights while the atomic-coordinate derivative acts on the electronic eigenvalues. This treatment may require further refinement when strong excitation introduces substantial electronic-structure renormalization.

The differentiability of the neural-network Hamiltonian is key to both the photoinduced force construction and the computational efficiency of TDAP-eML. Once the MLFF and ML Hamiltonian are trained, the geometry-dependent electronic structure and its coordinate derivatives can be evaluated through neural-network inference and automatic differentiation, avoiding repeated direct TDDFT calculations along extended trajectories. This leads to an increasing computational advantage with system size, approaching three orders of magnitude reduction in computer load for the larger systems considered here. More broadly, recent advances in ML electronic-structure models have demonstrated access to hybrid-functional accuracy, many-body quasiparticle corrections, and Hamiltonians incorporating spin--orbit couplings~\cite{tang2024deephybrid,hou2024manybodyml,qian2026maceh}. Incorporating such higher-accuracy and richer electronic descriptions into TDAP-eML could extend photoexcitation-induced lattice simulations beyond the accuracy and scale readily accessible to standard first-principles propagation such as TDDFT.

Taken together, these results establish TDAP-eML as a scalable framework linking nonequilibrium electronic evolution, photoinduced forces, and coherent lattice dynamics. Across prototype examples including Si and FeSe, TDAP-eML reproduces key first-principles features of photoexcited energy landscapes and structural dynamics, including polarization- and fluence-dependent coherent phonon generation and transient optical spectroscopy. Its favorable scaling and flexible ML electronic representations extend this capability to larger systems, longer time scales, and potentially higher-level electronic-structure descriptions. TDAP-eML therefore opens an unprecedented route to predictive simulations of light-driven structural dynamics beyond the spatial, temporal, and precision 
regimes accessible to direct first-principles propagation.

\section{Methods}

\subsection{Machine-learned force fields and electronic Hamiltonian}

TDAP-eML is implemented as a Python workflow integrating GPTFF~\cite{xie2024gptff}, HamGNN~\cite{zhong2023transferable} and the Atomic Simulation Environment (ASE)~\cite{larsen2017atomic}. GPTFF serves as the MLFF and provides the reference ground-state PES, $E_g(\mathcal{R})$, and the corresponding force contribution, $\mathbf{F}_i^g$. HamGNN provides the geometry-dependent ML Hamiltonian, $\hat{H}_{\mathbf{k}}(\mathcal{R})$, from which the instantaneous electronic structure is evaluated for each atomic configuration.

Because the localized atomic-orbital basis is nonorthogonal, the generalized eigenvalue problem is solved using both $\hat{H}_{\mathbf{k}}(\mathcal{R})$ and the overlap matrix $\hat{S}_{\mathbf{k}}(\mathcal{R})$, yielding the eigenvalues $\varepsilon_{n\mathbf{k}}(\mathcal{R})$ and eigenstates $\psi_{n\mathbf{k}}(\mathcal{R})$ entering the occupation-dependent electronic energy and force contributions. The overlap matrix and its atomic-coordinate derivatives, $\partial \hat{S}_{\mathbf{k}}/\partial \mathbf{R}_i$, required for the Pulay term are evaluated from the localized atomic-orbital basis following the OpenMX formalism~\cite{ozaki2003variational}. Their evaluation does not require an additional self-consistent electronic-structure calculation.

At each MD step, the electronic quantities are evaluated for the instantaneous atomic configuration and combined with the MLFF contribution to obtain the total force governing the photoexcited lattice dynamics. ASE then propagates the atomic coordinates using a velocity Verlet integrator. Unless otherwise stated, the trajectories are evolved without a thermostat or barostat. When lattice damping is included, the damping force defined below is added explicitly to the equations of motion.

\subsection{Electronic population propagation and relaxation}

The initial nonequilibrium population change $\Delta f_{n\mathbf{k}}$ is obtained from the post-pulse TDDFT calculation. For the PES benchmarks, this distribution is retained while the atomic structure is displaced along the selected phonon coordinates, allowing the structural dependence of the photoexcited PES to be evaluated under the same excitation condition. For the MD simulations, the post-pulse distribution provides the initial occupations for the subsequent electronic evolution.
During MD, the instantaneous electronic states are updated with the evolving atomic configuration, and the electronic amplitudes are propagated according to Eq.~\eqref{eq:a_discrete}. The resulting $\Delta f_{n\mathbf{k}}(t)$ is used at each MD step to update the effective PES and photoinduced forces through Eqs.~\eqref{eq:excited_pes} and \eqref{eq:excited_force_correction}.

Electronic relaxation is included phenomenologically through
\begin{equation}
\Delta f_{n\mathbf{k}}(t+\Delta t)
\rightarrow
\Delta f_{n\mathbf{k}}(t+\Delta t)
\exp\left(-\frac{\Delta t}{\tau_e}\right),
\end{equation}
where $\tau_e$ is an effective electronic relaxation time controlling the decay of the nonequilibrium population change.

When lattice damping is included, an additional friction force
\begin{equation}
\mathbf{F}_i^{\mathrm{damp}}
=
-\gamma_i M_i\dot{\mathbf{R}}_i
\end{equation}
is applied to the nuclear dynamics, where $\gamma_i$ is related to the coherent-phonon relaxation time by $\tau_{\mathrm{ph},i}=2/\gamma_i$ in the underdamped regime. The electronic and phonon relaxation times are set to $\tau_e=260~\mathrm{fs}$ and $\tau_{\mathrm{ph}}=1.3~\mathrm{ps}$ in the present case, respectively, consistent with characteristic relaxation times reported for photoexcited carriers and coherent phonons~\cite{sabbah2002femtosecond,hase2003birth}.

\subsection{Optical-response calculations}
The transient absorption spectrum is evaluated within the independent-particle approximation using linear-response theory~\cite{adler1962quantum,wiser1963dielectric} for each atomic configuration along the TDAP-eML trajectory. For a given MD snapshot $\mathcal{R}(t)$, the learned Hamiltonian is diagonalized to obtain the eigenvalues $\varepsilon_{n\mathbf{k}}(\mathcal{R}(t))$ and eigenstates $|n\mathbf{k}\rangle$. In this work, we focus on the absorption response along the $x$ direction, which is evaluated as
\begin{equation}
A_x(\omega,t)
\propto
\frac{1}{\hbar\omega}
\sum_{\mathbf{k},n,m}
w_{\mathbf{k}}
\left[
f_{n\mathbf{k}}(t)-f_{m\mathbf{k}}(t)
\right]
\left|
\langle m\mathbf{k}|\hat{v}_x|n\mathbf{k}\rangle
\right|^2
G_\eta
\left[
\varepsilon_{m\mathbf{k}}(t)-\varepsilon_{n\mathbf{k}}(t)-\hbar\omega
\right],
\end{equation}
where $\hbar\omega$ is the probe-photon energy, $f_{n\mathbf{k}}(t)$ is the instantaneous electronic occupation, $\hat{v}_x$ is the velocity operator along the $x$ direction, and $G_\eta$ is a Gaussian broadening function with width $\eta$.
%The velocity matrix element is evaluated from the predicted Hamiltonian, with $\hat{v}_x=(1/\hbar)\partial \hat{H}_{\mathbf{k}}/\partial k_x$. The transient absorption signal is obtained from the change in $A_x(\omega,t)$ relative to the reference spectrum before lattice motion.

\subsection{Datasets and simulation settings}

First-principles training datasets are generated using OpenMX within the Perdew--Burke--Ernzerhof (PBE) generalized-gradient approximation~\cite{perdew1996generalized}. For each material, 2,000 configurations are sampled by perturbing the relaxed equilibrium structure over the structural range relevant to the photoinduced lattice dynamics. Total energies and atomic forces are used to train the GPTFF model, while the corresponding localized atomic-orbital Hamiltonian and overlap information are used to train the ML Hamiltonian model. Details of the dataset construction and model training are provided in Supplementary Note~1 and Supplementary Table~1. The corresponding model benchmarks are presented in Supplementary Note~2 and Supplementary Figs.~1 and 2.

Reference TDDFT-MD calculations are performed using the time-dependent plane-wave (TDPW) package~\cite{meng2008real,lian2018photoexcitation}. The same PBE generalized-gradient approximation is used together with norm-conserving pseudopotentials and a plane-wave cutoff of $100~\mathrm{Ry}$. The Brillouin zone is sampled using $\Gamma$-centered $6\times6\times6$ and $10\times10\times6$ $k$-point meshes for Si and FeSe, respectively. These calculations provide the reference electronic excitation, photoexcited PESs and lattice dynamics used to benchmark TDAP-eML.

Photoexcitation in the TDPW simulations is driven by a Gaussian-envelope pump electric field,
\begin{equation}
\mathbf{E}(t)
=
E_0\hat{\mathbf{e}}
\cos\left[\omega(t-t_0)\right]
\exp\left[-\frac{(t-t_0)^2}{2\sigma^2}\right],
\end{equation}
where $E_0$ is the field amplitude, $\hat{\mathbf{e}}$ is the polarization direction, $\omega$ is the carrier frequency, $t_0$ is the pulse center and $\sigma$ characterizes the temporal width of the field envelope. The pump photon energies are $2.6~\mathrm{eV}$ for Si and $1.5~\mathrm{eV}$ for FeSe, with $\sigma=6~\mathrm{fs}$. A time step of $0.1~\mathrm{fs}$ is used for both the TDPW and TDAP-eML simulations.

\backmatter

\bibliography{bib}
\clearpage

\section{Data availability}
Data supporting the findings of this study are available from the corresponding author upon reasonable request.

\section{Code availability}
The code used to generate the results reported in this study will be made publicly available upon publication.

\section*{Author contributions}
L.C. and S.M. conceived the project and designed the research. L.C. developed the TDAP-eML framework, performed the first-principles and machine-learning calculations, and carried out the data analysis. Y.J. contributed to scientific discussions. D.C. provided technical support for the TDDFT calculations. S.M. supervised the research. L.C. and S.M. wrote the manuscript. All authors discussed the results and contributed to the final manuscript.

\section*{Competing interests}

The authors declare no competing interests.

\end{document}